\documentclass{article}
\usepackage{amssymb,color,graphicx}
\usepackage{quant}

\newcommand{\uref}[1]{(\ref{#1})}
\title{Classification of quantum channels of information transfer}

\author{Constantin V. Usenko}

\begin{document}

\maketitle

\abstract{	Classification of states of quantum channels of information transfer is built on the basis of unreducible representations of qubit state space group of symmetry and properties of density matrix spectrum. It is shown that pure disentangled states form two-dimensional surface, and the reason of state disentanglement is in degeneration of non-zero density matrix eigenvalues.
}

\section{ Group of symmetry  of Quantum Information}
Group of symmetry in quantum information theory \(\mathcal{Q}\) is the group of unitary transformations of qubit state space \(U(2)\) (see \cite{Vaccaro,Man}).

In reality important is Lie algebra of quantum information \(\mathcal{AQ}=u(2)\) generated by operators \(J_3\),\(J_+\),\(J_-\):
\begin{equation}
	\left[J_-J_+\right]=2J_3;\ \left[J_3J_\pm\right]=\pm J_\pm. 
\end{equation}
The report deals with physically substantial consequences of existence of group of symmetry of quantum information.\\
	In problems of quantum information one applies term qubit to arbitrary two-level quantum system. Qubit pure state is characterized by state vector -- normalized vector $\cat{\psi}$:
\begin{equation}
	 \cat{\psi}=\cos{\frac{\theta}{2}}\cat{0}+e^{i\phi}\sin{\frac{\theta}{2}}\cat{1}.
\end{equation}
 
 Set of state vectors forms a sphere -- Poincare sphere in quantum optics, or Bloch sphere in laser physics. Group of unitary transformations \(U(2)\) is the group of symmetry of qubit sphere.
\begin{figure}[h]
\begin{center}
	\includegraphics[scale=.8]{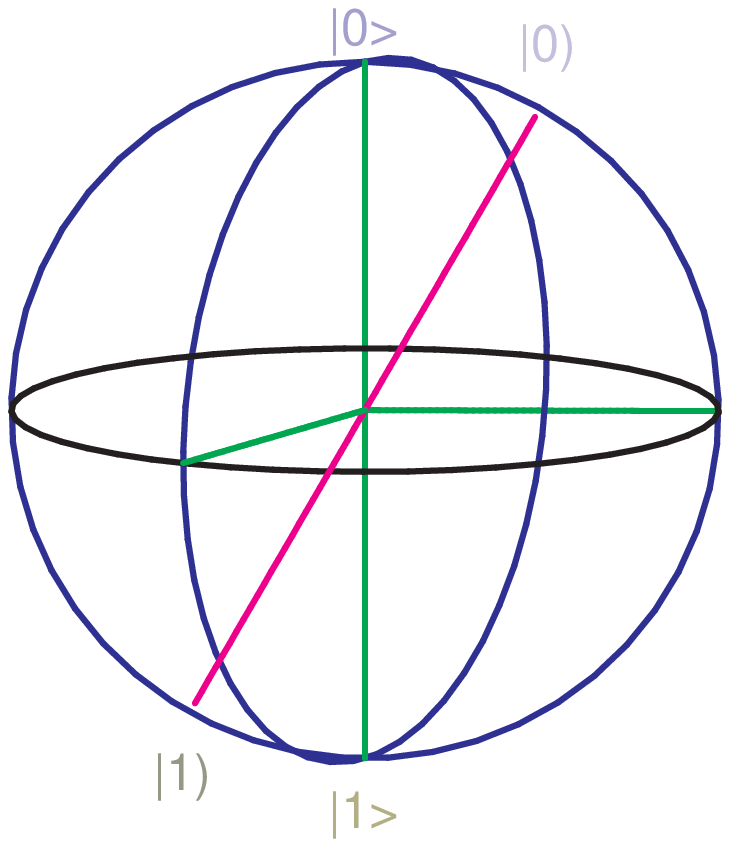}
	\caption{Space of qubit state vectors: \(\cat{0,1}\) and \(\catr{0,1}\)  are two bases.}
\end{center}
	\label{fig:1}
\end{figure}
 
\section{Information transfer channels}
\paragraph{Information} is obtained through comparison of results of measurements to expected results \cite{MP,Berry}. In the case of given probability distribution $p_k$ for each of $k$ variants of expected results of measurement, value of information obtained together  with expected result  $m$, is \(I_m\), and average value of information obtained in each sequential measurement is determined by Shannon entropy $S_S$:
\begin{equation}
	I_m\stackrel{def}{=}-\log p_m;\ S_S\stackrel{def}{=}-\sum{p_m\log p_m}.
\end{equation}
\paragraph{Information transfer channel} is an arbitrary 
device used for information processing (even in the case it just stores it during given time). Channel is prepared by means of source in given state, after that the state of the channel is measured by detector. Taken together, source + channel + detector are called channel as well. Sequence of states can be random or ordered, it forms mixed state. 
\begin{figure}[h]
\begin{center}
	\includegraphics[scale=.8,bb=0 0 2.2in 2.2in]{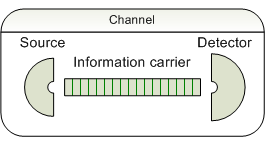}
	\caption{Information transfer channel and its components: Source, Media and Detector.}
\end{center}
	\label{fig:2}
\end{figure}
Ordered sequence of pure states is used for deliberate information transfer.

\paragraph{Quantum channel} of information transfer differs by fact that processes of channel state preparation, state distribution and its measurement can be explained with use of quantum physics laws only -- laws of classical physics are not enough for at least one of those three processes.

\subsection{Parallel channels}

Set of all the expected results of measurements is finite, that's why for information transfer use of devices with finite state number is just enough. 

In classical physics all states from finite set one can give in bitwise representation. One can for instance give numbers to points of state space and write down numbers of those in binary representation
\(\left\{k=\left[b_0,\ldots,b_p\right]\right\}\).
 In such way decomposition of arbitrary classical information transfer channel to set of bits takes place. Each of bits can be transfered by separate sub-channel (parallel channel), or one after another by sequential channel.
\subsubsection{Decomposition problem}
Quantum channel, in spite of classical one, instead of finite set of states has state space with finite dimensionality, that's why problem on decomposition becomes much more complicated \cite{Lew}.

Let us consider quantum channel with state space being product of two subspaces with dimensionalities $N_a$ and $N_b$. Set of state vectors of such channel is hyper-sphere with complex dimensionality $N_aN_b-1$. Division of channel into two subchannels with dimensionalities $N_a$ and $N_b$ gives as sets of pure states of subchannels hyperspheres with dimensionalities  $N_a-1$ and $N_b-1$. Direct product of those sets has dimensionality  $N_aN_b-N_a-N_b+1$ smaller than $N_aN_b-1$ -- not every pure state of quantum channel can be represented by pure states of subchannels. The states that have no such representation have got names of entangled states, for those total correlation (or anticorrelation) between results of measurements of subchannels is specific.
 
\subsection{Density matrix}
Set of channel states formed by source forms basis \(B=\left\{\cat{k};\ k=1\ldots N\right\}\) of state space -- finite-dimensional Hilbert space. 

Sequence of states generated by source can be characterized by one vector, in such case it forms pure state, or by set of different vectors, in such case it is a mix.

Universal method for description of states is given by density matrix. For pure state density matrix is projector to one-dimensional subspace of vectors collinear to state vector
\begin{equation}
	\hat{\rho}\left(\psi\right)=\cat{\psi}\otimes\bra{\psi},
\end{equation}
for mixed state density matrix is weighted mix of such projectors
\begin{equation}\label{rhodef}
	\hat{\rho}=\sum_{k=1}^N{p_k\roa{k}}; \ \forall k:\; 0\leq p_k\leq1;\ \sum_{k=1}^N{p_k}=1.
\end{equation}
Eigenvalues of density matrix are determined by probabilities or relative frequencies of production of respective eigenstates, and eigenvectors are the pure states produced by channel source.

Irrespectively of fact if sequence of states is random or ordered density matrix remains the same, sorting in process  of state production by source affects only time correlations between separate events of information transfer. Methods of investigation of information transfer channel states are equal in effectiveness in cases of spontaneous production of states and in problems on information transfer as well, till those do not deal with time correlations.

\subsection{Multi-state quantum channels of information transfer}
State space of classical channel of information transfer is finite and it can always be represented  as direct product of state spaces of two-state subchannels -- for classical information transfer channel there exists possibility of decomposition of channel to set of bits. Consequences of this fact are present in all the discrete mathematics -- mathematical apparatus of classical information theory.

As to quantum channels, main consequence of existence of entangled states is in unsolvability of problem on decomposition of quantum channel of information transfer to subchannels with smaller dimensionality.

Density matrix as Hermitian matrix with unit spur has $N^2-1$ independent real parameters. For multi-state quantum channel of information transfer with dimensionality of state space  $N_aN_b$, that is product of dimensionalities of state spaces of subchannels $N_a$ and $N_b\leq N_a$,  number of free parameters $\left(N_a^2-1\right)\left(N_b^2-1\right)=N^2+1-N_a^2-N_b^2$ is smaller than needed one. Thus properties of channel state are not limited by properties of states of subchannels -- additional parameters of channel state are needed (for instance, correlation coefficients).
\subsubsection{Induced and entangled bases}

Hereinafter we consider paired quantum channel of information transfer in which it is possible to register simultaneously states of two subchannels (source of states prepares one common channel state).

Induced basis of common state space $\left\{\catt{k}\in  \mathcal{H}\right\}$ is represented by direct products of bases $\left\{\cat{m}\in  \mathcal{H}_A\right\}$ of state space of particle of sort $A$  and $\left\{\catr{n}\in  \mathcal{H}_B\right\}$ of sort $B$ \begin{equation}
	\catt{k=m\cdot n}=\cat{m}\otimes\catr{n}\  \forall m,n.
\end{equation}
\begin{figure}[h]
\begin{center}
	\includegraphics[bb=0 0 379 190, scale=0.6]{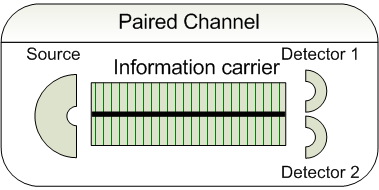}
	\caption{Paired channel and its components: channel has common source and two different detectors.}
\end{center}
	\label{fig:3}
\end{figure}

Common basis is result of arbitrary unitary transformation of induced basis. In the case of such transformation not being product of unitary transformations of subspaces at least part of vectors of common basis is some linear combination of direct products and corresponds to entangled states.
\begin{equation}\label{arbasis}
	\catt{k}=\sum_{\forall n}{c^{\left\{k\right\}}_{n}\cat{\mu_k\left(n\right)}\otimes\catr{n}}
\end{equation}
Such entangled basis is useful for instance in the case of its vectors being eigenvectors of density matrix. It is characterized with set of coefficients  \(c^{\left\{k\right\}}_{n}\) and with sequence of numbers \(\mu_k\left(n\right)\).

\subsubsection{Density matrix of paired channel}

Own basis of density matrix \uref{rhodef} is usually formed by vectors of state space of paired channel that are some non-trivial combinations \uref{arbasis} of vectors of induced basis, and representation of density matrix in induced basis has non-diagonal components
\begin{equation}\label{red_rho}
	\hat{\rho}=
		\sum_{\forall n,n'}{\Big[\sum_{\forall k}{p_{k}c^{\left\{k\right\}}_{n} c^{*\left\{k\right\}}_{n'}}\Big]
	\left(\cat{\mu_k\left(n\right)}\otimes\bra{\mu_k\left(n'\right)}\right)\otimes
	\left(\catr{n}\otimes\brar{n'}\right)}.
\end{equation}
\subsubsection{Density matrices of subchannels}
Density matrix of subchannel is obtained through averaging density matrix of paired channel by states of second subchannel 
\[
	\hat{\rho}^{\left\{a\right\}}=\sum_{\forall n}{\brar{n}\hat{\rho}\catr{n}}=\sum_{\forall m,m'}{\sum_{\forall n}{\rho_{m,m';n,n}}\cat{m}\bra{m'}}\]
	\[
	\hat{\rho}^{\left\{b\right\}}=\sum_{\forall m}{\bra{m}\hat{\rho}\cat{m}}
=\sum_{\forall n,n'}{\sum_{\forall m}{\rho_{m,m;n,n'}}\catr{n}\brar{n'}}
\]

With account of explicit form of density matrix in induced basis \uref{red_rho} we have
\begin{equation}\label{rho_ab}
	\hat{\rho}^{\left\{a\right\}}=
	\sum_{\forall m}{p_{m}^{\left\{a\right\}}\roa{m}};\  
	\hat{\rho}^{\left\{b\right\}}=
\sum_{\forall n}{p_{n}^{\left\{b\right\}}\eroa{n}}
\end{equation}
\begin{equation}\label{p_ab}
	p_{m}^{\left\{a\right\}}=
	\sum_{\forall k \forall n}{p_{k}\left|c^{\left\{k\right\}}_{n}\right|^2\delta_{m,\mu_k\left(n\right)}};\  
	p_{n}^{\left\{b\right\}}=
	\sum_{\forall k }{p_{k}\left|c^{\left\{k\right\}}_{n}\right|^2}. 
\end{equation}

\subsection{Channel types}
Depending on properties of own basis density matrix of paired channel in specific cases can be a product of density matrices of subchannels, or a mix of such products -- both those cases can take place for classical paired information transfer channel as well.  

Quantum properties of paired channel, particularly ones without classical analog, take place in all the cases when channel state can not be reduced to some classical variant, like entangled states of paraqubit can not be reduced to classical analog. Non-classical states, similarly to case of paraqubit, are given name of entangled states, that's why all the states of paired quantum channel of information transfer belong to one of the following three types:
\begin{enumerate}
	\item Independent subchannels: $\hat{\rho}=\hat{\rho}_A\otimes\hat{\rho}_B$;
	\item Mix of independent subchannels: $\hat{\rho}=  \sum_{s}{p_s\hat{\rho}^{\left(s\right)}_A \otimes\hat{\rho}^{\left(s\right)}_B}$;
	\item Entangled states of subchannels: all the others.
\end{enumerate}
Negative definition of entanglement leads to need in search of characteristics and criteria of entanglement. Further in this report results of symmetry analysis of entanglement of states of paired quantum channel of information transfer are given. 

\section{Representation of Lie algebra of quantum information in terms of channel state space}

Set of bases of \(N\)-dimensional state space of quantum channel is the orbit of group \(U(N)\) of unitary transformations of the space. Each unitary transformation of basis \(\cat{l}=\sum{U(l,k)\cat{k}}\) leaves density matrix the same and changes coefficients, \[		\rho_{l,l'}=\sum{U(l,k)\rho_{k,k'}U^+(l',k')}.\]
This expression gives representation of group of symmetry of quantum information in state space of information transfer quantum channel with arbitrary finite dimensionality.\\

\subsection{Ladder operators}

For each set of basis vectors, thus for each quantum information transfer channel  there exists (by construction, see \cite{cvu1}) system of operators 
\begin{equation}\label{ladder_def}
\hat{J}_+\stackrel{Def}{=} \sum^{N}_{k=1}{\sqrt{\left(N-k\right)k}\cat{k+1}\otimes\bra{k}};\ 
 \hat{J}_-\stackrel{Def}{=} \sum^{N}_{k=1}{\sqrt{\left(N+1-k\right)\left(k-1\right)}\cat{k-1}\otimes\bra{k}};
\end{equation}
\[
 \hat{J}_3\stackrel{Def}{=} \big(\hat{J}_-\hat{J}_+-\hat{J}_+\hat{J}_-\big)/2
 =\sum^{N}_{k=1}{\left(k-(1+N)/2\right)\cat{k}\otimes\bra{k}}.
\]
Result of effect of operator $\hat{J}_{+}$ on arbitrary vector $\cat{k}$ is in following vector $\cat{k+1}$ or zero (in the case of vector with the largest number).  Operator \(\hat{J}_3\)
has basis vectors as eigenvectors. Re-denotation $\cat{k}\rightarrow\cat{m=k-(1+N)/2}: m=-N/2\ldots N/2$ completes analogy between arbitrary basis and basis of irreducible representation of group \(U(2)\). The group \(U(2)\) can be represented by orthogonal transformations of ladder operators as well.

\subsection{Representation of density matrix}

Ladder operators realize irreducible representation of Lie algebra of group of invariance of quantum information. In the case if from physical considerations it is needed to change at least sequence of basis vectors one has to change simultaneously ladder operators of basis -- ladder operators are associated to basis of state space. From the other hand, ladder operators are generators of algebra of operators in the meaning that arbitrary operator has representation by double series 
\begin{equation}
	\hat{A}=\sum{A_{k,l}\hat{J}_{+}^k\hat{J}_{-}^l}.
\end{equation}

Eigenvectors of density matrix form basis to which system of ladder operators corresponds \uref{ladder_def}. In this basis it is possible to represent by means of some interpolating function 
\begin{equation}
	p\left(x\right):\ p\left(k-\frac{1+N}{2}\right)=p_k, 
\end{equation}
density matrix in form invariant with respect to unitary transformations of state space:
\begin{equation}
	\rho=p\left(\vec{n}\cdot\vec{J}\right).
\end{equation}
Here denotation \(\vec{n}\cdot\vec{J}=n_3J_3+n_-J_++n_+J_-\) is used.

\subsection{Representation of pure states of paired channel}
Ladder operators of channel are expanded to operators of subchannels, similarly to division of operator of total moment $\hat{J}_a=\hat{L}_a+\hat{S}_a$  to spin $\hat{S}_a$, and orbital $\hat{L}_a$ ones. 

Result is in  
classification of pure states of paired channel
 by irreducible representations.

Let us denote  $l=\left(N_B-1\right)/2$ and $s=\left(N_A-1\right)/2 \le l$. Basis in state space of channel 
\begin{equation}
	\left\{\catt{j,m_j}; m_j=-j,\ldots j;j=l-s,l-s+1,\ldots l+s\right\},
\end{equation}
is formed by eigenvectors of operators  $\hat{J}_3$ and $\hat{J}^2= \hat{J}^2_3+\hat{J}_3+\hat{J}_+\hat{J}_-$: 
\begin{equation}
	\hat{J}_3\catt{j,m_j}=m_j\catt{j,m_j};\ \hat{J}^2\catt{j,m_j}=j(j+1)\catt{j,m_j}.
\end{equation}
Expansion of eigenvectors by induced basis
\begin{equation}\label{clebsh}
	\catt{j,m_l+m_s}=\sum{
	C_{j,m_l;m_s}\cat{l,m_l}\otimes\catr{s,m_s},
	}
\end{equation}
is given by Clebsch-Gordan coefficients
\[C_{j,m_l;m_s}=\sum_k{\frac{(-1)^k}{k!(l-m_l-k)!(s+m_s-k)!(l+s-j-k)!(j-l+m_s+k)!(j-s-m_l+k)!}}.
\]
In this basis only two limiting states
\begin{equation}\label{dist_state}
	\catt{l+s,\pm\left(l+s\right)}=\cat{l,\pm l}\otimes\catr{s,\pm s},
\end{equation}
are products of one-particle ones, all the others correspond to entangled states.

\subsection{Representation of mixed states of paired channel}
Mixed states of paired quantum channel of information transfer are weighted mix of states produced by source, that's why density matrices of those are diagonal in basis of source states
\[
	\hat{\rho}=
	\sum_{j=l-s}^{l+s}{\sum_{m=-j}^{j}{p_{j,m}\proa{j,m}	}};\ 
	\sum_{j=l-s}^{l+s}{\sum_{m=-j}^{j}{p_{j,m}	}}=1.
\]
In induced basis density matrix of paired channel has non-zero non-diagonal components 
\begin{equation}\label{full}
	\hat{\rho}=\sum_{j=l-s}^{l+s}{\sum_{m=-j}^{j}{p_{j,m}\sum_{k,n=-s}^{ s}{C_{j,m-k;k}C_{j,m-n;n}
	\cat{m-k}\bra{m-n}\otimes\ecat{k}\ebra{n}}}
	}.
\end{equation}
Density matrices of subchannels according to \uref{rho_ab} are diagonal:
\begin{equation}
	\hat{\rho}^{\left\{a\right\}}=
	\sum_{\forall m}{p_{m}^{\left\{a\right\}}\roa{m}};\  
	\hat{\rho}^{\left\{b\right\}}=
\sum_{\forall n}{p_{n}^{\left\{b\right\}}\eroa{n}}.
\end{equation}
 Probabilities of states of subchannels  \uref{p_ab} are
\begin{equation}
	p_k^{\left\{a\right\}}=
	\sum_{j=l-s}^{l+s}{\sum_{n=-s}^{s}{p_{j,k+n}C_{j,k;n}^2}	
	};\
		p_n^{\left\{b\right\}}=\sum_{j=l-s}^{l+s}{\sum_{k=-l}^{ l}{p_{j,n+k}C_{j,k;n}^2}}.
\end{equation}
Probabilities of simultaneous registration of one subchannel in state $k$ and the other one in state $n$ are
\begin{equation}
	P_{k,n}=\sum_{j=l-s}^{l+s}{p_{j,n+k}C_{j,k;n}^2 }.
\end{equation}
Full correlation between states of subchannels takes place in specific cases, like pure states, only. From the other hand, loss of correlation between states of subchannels is possible in specific cases of reduction of coefficients at non-diagonal elements of common density matrix in induced basis \uref{full} only.

 Main conclusion is in exceptionality of not entangled states.

\section{Disentanglement of states}
Reasons for absence of entanglement in typical states of classical channels of information transfer are:
\begin{itemize}
	\item 
Disentanglement of states can come to existence due to degeneration of eigenvalues of density matrix.

	\item 
Totally degenerated density matrix is proportional to unit matrix that is product of unit matrices of subsystems, that's why common density matrix is product of density matrices of subsystems, thus it corresponds to independent subchannels of information transfer. 

	\item 
Degeneration is specific to each pure state in which density matrix is  $N-1$-times degenerated  ($\rho_k^{deg}=0$) as well, though among pure states only two correspond to independent subchannels of information transfer. 
\end{itemize}

Thus for some pairs of eigenvalues only result of degeneration of density matrix is in disentanglement of state.  

\subsection{Disentanglement of paraqubit}

The simpliest example of two-particle quantum channel of information transfer -- paraqubit -- is a rather clear example of effect of degeneration of density matrix on disentanglement of states.

\subsubsection{Basis of irreducible representations of paraqubit}

Two irreducible representations are singlet $j=1/2-1/2=0$:
\[
	\catt{s}=\frac{1}{\sqrt{2}}\cat{\frac{1}{2}}\otimes\ecat{-\frac{1}{2}}
	-\frac{1}{\sqrt{2}}\cat{-\frac{1}{2}}\otimes\ecat{\frac{1}{2}};
\]
 and triplet $j=1/2+1/2=1$:  
\[
\begin{array}{ll}
	\catt{d}&=\cat{-\frac{1}{2}}\otimes\ecat{-\frac{1}{2}};\\
	 \catt{0}&=\frac{1}{\sqrt{2}}\cat{\frac{1}{2}}\otimes\ecat{-\frac{1}{2}}
	+\frac{1}{\sqrt{2}}\cat{-\frac{1}{2}}\otimes\ecat{\frac{1}{2}};\\
	\catt{u}&=\cat{\frac{1}{2}}\otimes\ecat{\frac{1}{2}}
\end{array}
\]
ones.
Arbitrary linear combination of vectors of triplet state
\(a\catt{d}+b \catt{0}+c\catt{u}=\cat{\psi}\otimes\ecat{\phi}\) 
is product of two vectors under condition 
\(a=k^2c;\ b=\sqrt{2}kc;\ c=\frac{1}{1+\abs{k}^2}
\).

Set of disentangled triplet states is topologically equivalent to sphere.

Fully disentangled basis is constructed by both subspaces of irreducible representations.

General state of paraqubit is 
\begin{equation}\label{full_pq}
	\hat{\rho}=p_s	\hat{\rho}_s + p_{d}\proa{d}+ p_{u}\proa{u}+ p_{0}\proa{0}
\end{equation}
\begin{itemize}
	\item $p_s$ -- part of singlet state;
	\item $p_{0}$ -- part of entangled triplet state;
	\item $p_{d}$, $p_{u}$ -- parts of states of independent particles.
\end{itemize}
In induced basis 
\begin{equation}\label{full_pqs}
\begin{array}{l}
	\hat{\rho}=	p_{d}\roa{0}\otimes\eroa{0}+p_{u}\roa{1}\otimes\eroa{1}\\
	 +	\frac{p_{0}+p_s}{2}\big(\roa{0}\otimes\eroa{1}+\roa{1}\otimes\eroa{0}\big)\\
+	\frac{p_{0}-p_s}{2}\big(
\cat{0}\bra{1}\otimes\ecat{1}\ebra{0}+
\cat{1}\bra{0}\otimes\ecat{0}\ebra{1}
\big)
\end{array},
\end{equation}
expression for density matrix has three types of terms. In the first line terms corresponding to classical state doubling are gathered. The second line consists of terms responsible for classical mix. Only in the third line two terms responsible for entanglement are gathered. Those terms disappear at coincidence of singlet and entangled triplet states $p_s=p_{0}$.

Coincidence of other pairs of eigenvalues does not generate disentanglement of states. 
 
\section{Conclusions}
\begin{itemize}
	\item Symmetry of state space of quantum channels of information transfer is determined by group \(U(2)\) of unitary transformations of qubit state spaces.  

	\item Quasi-classical, i.e. not entangled states of quantum channel of information transfer with arbitrary dimensionality form sphere isomorphous to Poincare sphere of pure qubit state.
	
\item Entangled states have as state vectors basis vectors of irreducible representations of group of symmetry of state space of information transfer quantum channels. Inputs of states of subchannels to each entangled state are determined by Clebsch-Gordan coefficients.

	\item Mixed states of quantum channel of information transfer are entangled. Those are disentangled in channels with degenerated density matrix.

\end{itemize}

\end{document}